\documentclass[npg]{copernicus}
\usepackage[bookmarks=true]{hyperref}
\usepackage[pdftex]{graphicx}
\usepackage{amsmath,amsfonts,amssymb}
\usepackage{algorithm}
\usepackage{algorithmic}
\usepackage{array}
\usepackage[tight,footnotesize]{subfigure}
\usepackage{url}
\usepackage{fancybox,color}
\usepackage{ulem,bm}

\def\E{\mathrm{E}}
\def\A{\mathbf{A}}

\newtheorem{theorem}{Theorem}
\begin{document}

\linenumbers

\title{DISTRIBUTED ALLOCATION OF MOBILE SENSING SWARMS IN GYRE FLOWS}

\author[1]{KENNETH MALLORY and M. ANI HSIEH}
\author[2]{ERIC FORGOSTON}
\author[3]{IRA B. SCHWARTZ}

\affil[1]{SAS LAB, DREXEL UNIVERSITY, PHILADELPHIA, PA 19104, USA}
\affil[2]{DEPARTMENT OF MATHEMATICAL SCIENCES, MONTCLAIR STATE UNIVERSITY, MONTCLAIR, NJ 07043, USA}
\affil[3]{NONLINEAR SYSTEMS DYNAMICS SECTION, PLASMA PHYSICS DIVISION, CODE 6792, U.S. NAVAL RESEARCH LAB, WASHINGTON, DC 20375, USA}


\runningtitle{ALLOCATION OF SWARMS IN GYRE FLOWS}

\runningauthor{MALLORY ET. AL.}

\correspondence{KENNETH MALLORY\\ (km374@drexel.edu)}

\received{}
\pubdiscuss{} 
\revised{}
\accepted{}
\published{}


\firstpage{1}

\maketitle  

\begin{abstract}
We address the synthesis of distributed control policies to enable a swarm of
homogeneous mobile sensors to maintain a desired spatial distribution in a
geophysical flow environment, or workspace. In this article, we assume the mobile sensors (or
robots) have a ``map'' of the environment denoting the locations of the
Lagrangian coherent structures or LCS boundaries.  Using this information,
we design agent-level hybrid control policies that leverage the
surrounding fluid dynamics and inherent environmental noise to enable the team
to maintain a desired distribution in the workspace.  We discuss the stability properties of the ensemble dynamics of the distributed control
policies.  Since realistic quasi-geostrophic ocean models predict double-gyre
flow solutions, we use a wind-driven multi-gyre flow model to verify the
feasibility of the proposed distributed control strategy and compare the
proposed control strategy with a baseline deterministic allocation strategy.  Lastly, we validate the control strategy using actual flow data obtained by our coherent structure experimental testbed.
\end{abstract}


\introduction  
Geophysical flows are naturally stochastic and aperiodic, yet exhibit coherent
structure.  Coherent structures are of significant importance since knowledge
of them enables the prediction and estimation of the underlying geophysical
fluid dynamics.  In realistic ocean flows, these time-dependent coherent
structures, or Lagrangian coherent structures (LCS), are similar to
separatrices that divide the flow into dynamically distinct regions, and are essentially
extensions of stable and unstable manifolds to general time-dependent flows
\citep{ref:Haller2000}.  As such, they encode a great deal of global information
about the dynamics and transport of the fluidic environment. For two-dimensional (2D) flows, ridges of
locally maximal finite-time Lyapunov exponent (FTLE) \citep{ref:Shadden2005}
values correspond, to a good approximation (though see~\citep{hall11}), to Lagrangian
coherent structures. Details regarding the derivation of the FTLE can be found in the
  literature~\cite{hall00,hall01,hall02,ref:Shadden2005,leshma07,brawig09}.

Recent years have seen the use of autonomous underwater and surface vehicles
(AUVs and ASVs) for persistent monitoring of the ocean to study the
dynamics of various biological and physical phenomena, such as plankton
assemblages \citep{ref:Caron2008}, temperature and salinity profiles
\citep{lynch2008, ref:Wu2011, ref:Sydney2011}, and the onset of harmful algae
blooms \citep{ref:Zhang2007, ref:Chen2008, ref:Das2011}.  These studies have
mostly focused on the deployment of single, or small numbers of, AUVs working
in conjunction with a few stationary sensors and ASVs.  While data collection
strategies in these works are driven by the dynamics of the processes they
study,  most existing works treat the effect of the surrounding fluid as
  solely external disturbances \citep{ref:Das2011, ref:Williams2012}, largely
  because of our limited understanding of the complexities of ocean
  dynamics. Recently, LCS have been shown to coincide with optimal
  trajectories in the ocean which minimize the energy and the time needed to
  traverse from one point to another \citep{ref:Inanc2005,
    ref:Senatore2008}. And while recent works have begun to consider the
  dynamics of the surrounding fluid in the development of fuel efficient
  navigation strategies \citep{ref:Lolla2012, ref:DeVries2011}, they rely
  mostly on historical ocean flow data and do not employ knowledge of LCS
  boundaries.

A drawback to operating both active and passive sensors in time-dependent
and stochastic environments like the ocean is that the sensors will escape
from their monitoring region of interest with some finite probability. This
  is because the escape likelihood of any given sensor is not only a function
  of the unstable environmental dynamics and inherent noise, but also the amount of control effort available to the sensor.  Since the LCS are inherently unstable and denote regions of the flow where escape events occur with higher probability \citep{fbys11}, knowledge of the LCS are of paramount importance in maintaining a sensor in a particular monitoring region.

In order to maintain stable patterns in unstable flows, the objective of this work is to develop decentralized control policies for a team of autonomous underwater vehicles (AUVs) and/or mobile sensing resources to maintain a desired spatial distribution in a fluidic environment. Specifically, we devise agent-level control policies which allow individual AUVs to leverage the surrounding fluid dynamics and inherent environmental noise to navigate from one dynamically distinct region to another in the workspace. While our agent-level control policies are devised using {\it a priori} knowledge of manifold/coherent structure locations within the region of interest, execution of these control strategies by the individual robots is achieved using only information that can be obtained via local sensing and local communication with neighboring AUVs. As such, individual robots do not require information on the global dynamics of the surrounding fluid.  The result is a distributed allocation strategy that minimizes the overall control-effort employed by the team to maintain the desired spatial formation for environmental monitoring applications.

While this problem can be formulated as a multi-task (MT), single-robot (SR),
time-extended assignment (TA) problem \citep{ref:Gerkey04}, existing
approaches do not take into account the effects of fluid dynamics coupled with
the inherent environmental noise \citep{ref:Gerkey02, ref:Dias06, ref:Dahl06,
  ref:Hsieh08b, ref:Berman08}. The novelty of this work lies in the use of
nonlinear dynamical systems tools and recent results in LCS theory
 applied to collaborative robot tracking \citep{ref:Hsieh2012} to synthesize distributed control policies that enables AUVs to maintain a desired distribution in a fluidic environment.

The paper is structured as follows: We formulate the problem and outline key
assumptions in Section \ref{sec:probForm}.  The development of the distributed
control strategy is presented in Section \ref{sec:method} and its theoretical
properties are analyzed in Section \ref{sec:analysis}.  Section
\ref{sec:results} presents our simulation methodology, results, and discussion.  We end with conclusions and directions for future work in Section \ref{sec:future}.

\section{Problem Formulation}\label{sec:probForm}
Consider the deployment of $N$ mobile sensing resources (AUVs/ASVs) to monitor $M$ regions in the ocean.  The objective is to synthesize agent-level control policies that will enable the team to autonomously maintain a desired distribution across the $M$ regions in a dynamic and noisy fluidic environment.  We assume the following kinematic model for each AUV:
\begin{equation}\label{eq:kinematics}
\dot{\mathbf{q}}_k = \mathbf{u}_{k} + \mathbf{v}^{f}_{\mathbf{q}_k} \quad k \in \{1, \ldots,n\},
\end{equation}
where $\mathbf{q}_k = [x_k, \, y_k, \, z_k]^T$ denotes the vehicle's position, $\mathbf{u}_{k}$ denotes the $3 \times 1$ control input vector, and $\mathbf{v}^{f}_{\mathbf{q}_k}$ denotes the velocity of the fluid experienced/measured by the $k^{th}$ vehicle.

\begin{figure}
\centering
\subfigure[]{\includegraphics[width=0.45\linewidth]{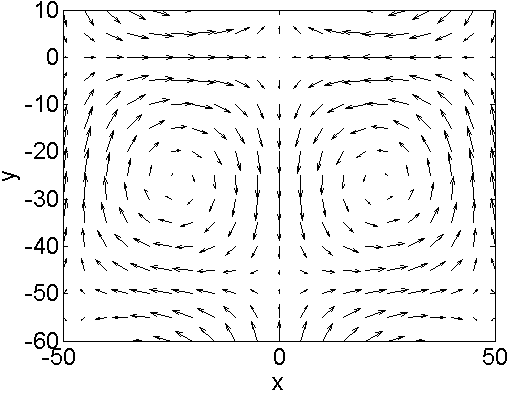}\label{fig:ppDbleGyre}}
\subfigure[]{\includegraphics[width=0.5\linewidth]{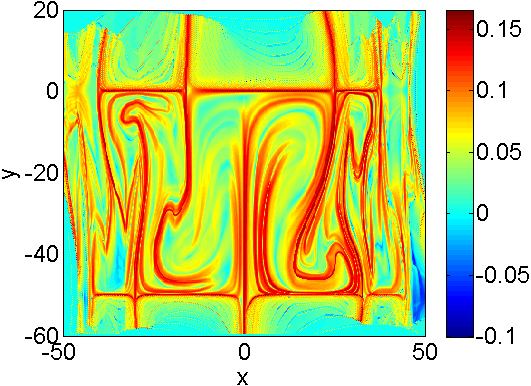}\label{fig:lcsDbleGyre}}
\caption{(a) Vector field and (b) FTLE field of the model given by \eqref{eq:doubleGyre} for two gyres with $A=
  10$, $\mu=0.005$, $\varepsilon = 0.1$, $\psi = 0$, $I = 0.01$, and $s = 50$. LCS are characterized by regions with maximum FTLE measures (denoted by red).  In 2D flows, regions with maximum FTLE measures correspond to 1D curves.}
\end{figure}

In this work, we limit our discussion to 2D planar flows and motions and thus we assume $z_k$ is constant for all $k$.  As such, $\mathbf{v}^{f}_{\mathbf{q}_k}$ is a sample of a 2D vector field denoted by $F(\mathbf{q})$ at $\mathbf{q}_k$ whose $z$ component is equal to zero, {\it i.e.}, $F_z(\mathbf{q}) = 0$, for all $\mathbf{q}$. Since realistic quasi-geostrophic ocean models exhibit multi-gyre flow solutions, we assume $F(\mathbf{q})$ is provided by the 2D wind-driven multi-gyre flow model given by
\begin{subequations}\label{eq:doubleGyre}
\begin{align}
&\dot{x}  = -\pi A \sin(\pi \frac{f(x,t)}{s}) \cos(\pi \frac{y}{s}) - \mu x + \eta_1(t), \\
&\dot{y}  = \pi A \cos(\pi \frac{f(x,t)}{s}) \sin(\pi \frac{y}{s}) \frac{df}{dx} - \mu y + \eta_2(t), \\
&\dot{z}  = 0,\\
&f(x,t)  = x + \varepsilon \sin(\pi \frac{x}{2s}) \sin(\omega t+ \psi). 
\end{align}
\end{subequations}
When $\varepsilon = 0$, the multi-gyre flow is time-independent, while for
$\varepsilon \neq 0$, the gyres undergo a periodic expansion and contraction
in the $x$ direction.  In~\eqref{eq:doubleGyre}, $A$ approximately
determines the amplitude of the velocity vectors, $\omega/2\pi$ gives the
oscillation frequency, $\varepsilon$ determines the amplitude of the
left-right motion of the separatrix between the gyres, $\psi$ is the phase,
$\mu$ determines the dissipation, $s$ scales the dimensions of the workspace,
and $\eta_i(t)$
describes a stochastic white noise with mean zero and standard deviation
$\sigma = \sqrt{2I}$, for noise intensity $I$.  Figures \ref{fig:ppDbleGyre}
and \ref{fig:lcsDbleGyre} show the vector field of a
two-gyre model and the corresponding FTLE curves for
  the time-dependent case.

\begin{figure}
\centering
\subfigure[]{\includegraphics[width=0.45\linewidth]{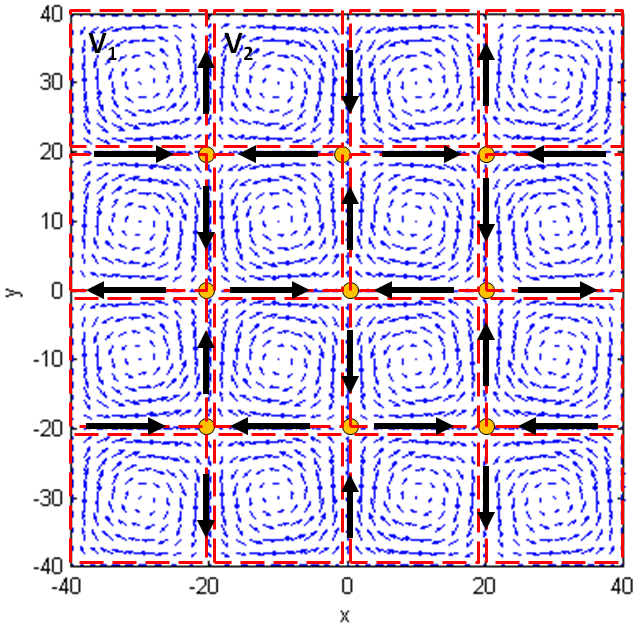}\label{fig:gyreTessellation}}
\subfigure[]{\includegraphics[width=0.45\linewidth]{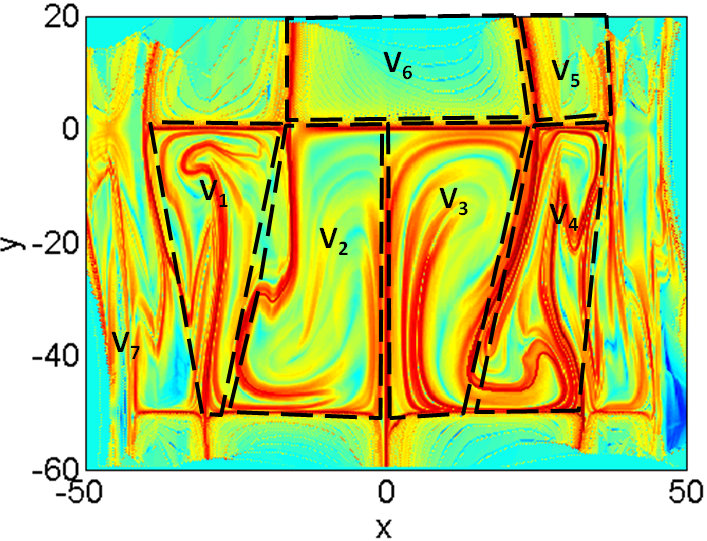}\label{fig:LCStessellation}}
\caption{Two examples of LCS-based cell decomposition of the region of interest assuming a flow field given by
  \eqref{eq:doubleGyre}. These cell decompositions were performed manually.  (a) A $4 \times 4$ time-independent grid of gyres
  with $A= 0.5$, $\mu=0.005$, $\varepsilon = 0$, $\psi = 0$, $I = 35$, and $s
  = 20$. The stable and unstable manifolds of each saddle point in the system
  is shown by the black arrows.  (b) An FTLE based cell decomposition for a time-dependent double-gyre system with the same parameters as Fig. \ref{fig:lcsDbleGyre}. \label{fig:tessellation}}
\end{figure}

Let ${\cal W}$ denote an obstacle-free workspace with flow dynamics given by
\eqref{eq:doubleGyre}.  We assume a tessellation of ${\cal W}$ such that the
boundaries of each cell roughly corresponds to the stable/unstable manifolds or
LCS curves quantified by maximum FTLE ridges as shown in
Fig. \ref{fig:tessellation}.  In general, it may be unreasonable to expect small
  resource constrained autonomous vehicles to be able to track the LCS locations in real time.  However, LCS boundary locations can be determined using historical data, ocean model data, {\it e.g.}, data provided by the Navy Coastal Ocean Model (NCOM) databases, and/or data obtained {\it a priori} using LCS tracking strategies similar to \citep{ref:Hsieh2012}.  This information can then be used to obtain an LCS-based cell decomposition of ${\cal W}$.
  Fig. \ref{fig:tessellation} shows two manual cell decompositions of the
  workspace where the cell boundaries roughly correspond to maximum FLTE
  ridges. In this work, we assume the tessellation of ${\cal W}$ is given and
  do not address the problem of automatic tessellation of the workspace to
  achieve a decomposition where cell boundaries correspond to LCS curves.

A tessellation of the workspace along boundaries characterized by maximum FTLE
ridges makes sense since they separate regions within the flow field that
exhibit distinct dynamic behavior and denote regions in the flow field where
more escape events may occur probabilistically \citep{fbys11}. In the
time-independent case, these boundaries correspond to stable and unstable
manifolds of saddle points in the system.  The manifolds can also be
characterized by maximum FTLE ridges where the FTLE is computed based on a
backward (attracting structures) or forward (repelling structures) integration
in time. Since the manifolds demarcate the basin boundaries separating the
distinct dynamical regions, they are also regions that are uncertain with
respect to velocity vectors within a neighborhood of the manifold. Therefore,
switching between regions in neighborhoods of the manifold is influenced both
by deterministic uncertainty as well as stochasticity due to external noise.

Given an FTLE-based cell decomposition of ${\cal W}$, let ${\cal G} = ({\cal
  V}, {\cal E})$ denote an undirected graph whose vertex set ${\cal V} =
\{V_1, \ldots, V_M\}$ represents the collection of FTLE-derived cells in
${\cal W}$.  An edge $e_{ij}$ exists in the set ${\cal E}$ if cells $V_i$ and
$V_j$ share a physical boundary or are physically adjacent.  In other words,
${\cal G}$ serves as a roadmap for ${\cal W}$.  For the case shown in
Fig. \ref{fig:gyreTessellation}, adjacency {of an interior cell} is defined
based on  four neighborhoods. Let $N_i$ denote the number of AUVs or mobile sensing resources/robots within $V_i$.  The objective is to synthesize agent-level control policies, or $\mathbf{u}_k$, to achieve and maintain a desired distribution of the $N$ agents across the $M$ regions, denoted by $\mathbf{\bar{N}} = [\bar{N}_1, \ldots, \bar{N}_M]^T$, in an environment whose dynamics are given by \eqref{eq:doubleGyre}.

We assume that robots are given a map of the environment, ${\cal G}$, and
  $\mathbf{\bar{N}}$.  Since the tessellation of ${\cal W}$ is given, the LCS
  locations corresponding to the boundaries of each $V_i$ is also known {\it a
    priori}.  Additionally, we assume robots co-located within the same $V_i$
  have the ability to communicate with each other.  This makes sense since
  coherent structures can act as transport barriers and prevent underwater
  acoustic wave propagation \citep{ref:Wang2009, ref:Rypina2011}.  Finally, we
  assume individual robots have the ability to localize within the workspace, {\it i.e.}, determine their own positions in the workspace.
These assumptions are necessary to enable the development of a prioritization
scheme within each $V_i$ based on an individual robot's escape likelihoods in
order to achieve the desired allocation.  The prioritization scheme will allow
robots to minimize the control effort expenditure as they move within the set
${\cal V}$.  We describe the methodology in the following section.

\section{Methodology}\label{sec:method}
We propose to leverage the environmental dynamics and the inherent
environmental noise to synthesize energy-efficient control policies for a team
of mobile sensing resources/robots to maintain the desired allocation in
${\cal W}$ at all times.  We assume each robot has a map of the environment.
In our case, this translates to providing the robots the locations
of LCS boundaries that define each $V_i$ in ${\cal G}$.  Since LCS curves separate ${\cal W}$ into regions with
distinct flow dynamics, this becomes analogous to providing autonomous ground or
aerial vehicles a map of the environment which is often obtained {\it a
  priori}.  In a fluidic environment, the map consists of the locations of the
maximum FTLE ridges computed from data and refined,
potentially in real-time, using a strategy similar to the one found in \citep{ref:Hsieh2012}.  Thus, we assume each robot has a map of the environment and has the ability to determine the direction it is moving in within the global coordinate frame, {\it i.e.}, the ability to localize.




\subsection{Controller Synthesis}
Consider a team of $N$ robots initially distributed across $M$ gyres/cells.  Since the objective is to achieve a desired allocation of $\mathbf{\bar{N}}$ at all times, the proposed strategy will consist of two phases: an auction to determine which robots within each $V_i$ should be tasked to leave/stay and an actuation phase where robots execute the appropriate leave/stay controller.

\subsubsection{Auction Phase}
The purpose of the auction phase is to determine whether $N_i(t) > \bar{N}_i$
and to assign the appropriate actuation strategy for each robot within
$V_i$. Let $Q_i$ denote an ordered set whose elements provide robot identities
 that are arranged from highest escape likelihoods to lowest escape likelihoods from $V_i$.

In general, to first order we assume a geometric measure whereby the escape likelihood of any particle within $V_i$ increases as it approaches the boundary of $V_i$, denoted as $\partial V_i$ \citep{fbys11}.
Given ${\cal W}$, with dynamics given by \eqref{eq:doubleGyre}, consider the
case when $\varepsilon = 0$ and $I \neq 0$, {\it i.e.}, the case when the
fluid dynamics is time-independent in the presence of noise.  The boundaries
between each $V_i$ are given by the stable and unstable manifolds of the
saddle points within ${\cal W}$ as shown in Fig. \ref{fig:gyreTessellation}.
While there exists a stable attractor in each $V_i$ {when $I=0$}, the presence
of noise means that robots originating in
$V_i$ have a non-zero probability of landing in a neighboring gyre $V_j$ where
$e_{ij} \in {\cal E}$.  Here, we assume that
robots experience the same escape likelihoods in each gyre/cell and assume
that  $P_k(\neg i|i)$, the probability that a robot escapes from region $i$
to an adjacent region,  can be estimated based on a
robot's proximity to a cell boundary with some assumption of the environmental
noise profile \citep{fbys11}.

Let $d(\mathbf{q}_k, \partial V_i)$ denote the distance between a robot $k$ located in $V_i$ and the boundary of $V_i$. We define the set $Q_i = \{k_1, \ldots, k_{N_i}\}$ such that $d(q_{k_1}, \partial V_i) \leq  d(q_{k_2}, \partial V_i) \leq \ldots \leq d(q_{N_i}, \partial V_i)$.  The set $Q_i$ provides the prioritization scheme for tasking robots within $V_i$ to leave if $N_i(t) > \bar{N}_i$.  The assumption is that robots with higher escape likelihoods are more likely to be ``pushed'' out of $V_i$ by the environment dynamics and will not have to exert as much control effort when moving to another cell, minimizing the overall control effort required by the team.

In general, a simple auction scheme can be used to determine $Q_i$ in a
distributed fashion by the robots in $V_i$ \citep{ref:Dias06}.  If $N_i(t) >
\bar{N}_i$, then the first $N_i - \bar{N}_i$ elements of $Q_i$, denoted by
$Q_{i_L} \subset Q_i$, are tasked to leave $V_i$.  The number of robots in
$V_i$ can be established in a distributed manner in a similar fashion.  The
auction can be executed periodically at some frequency $1/T_a$ where $T_a$
denotes the length of time between each auction and should be greater than the relaxation time of the AUV/ASV dynamics.

\subsubsection{Actuation Phase}
\begin{figure}
\centering
\includegraphics[width=0.75\linewidth]{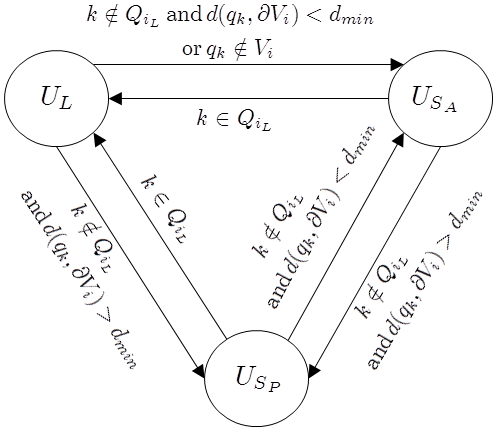}
\caption{Schematic of the single-robot hybrid robot control policy.\label{fig:controller}}
\end{figure}

For the actuation phase, individual robots execute their assigned controllers
depending on whether they were tasked to stay or leave during the auction
phase.  As such, the individual robot control strategy is a hybrid control
policy consisting of three discrete states: a {\tt leave} state, $U_L$, a {\tt stay} state, $U_S$, which is further distinguished into $U_{S_A}$ and $U_{S_P}$.  Robots who are tasked to {\tt leave} will execute $U_L$ until they have left $V_i$ or until they have been once again tasked to {\tt stay}. Robots who are tasked to {\tt stay} will execute $U_{S_P}$ if $d(q_{k}, \partial V_i) > d_{min}$ and $U_{S_A}$ otherwise.  In other words, if a robot's distance to the cell boundary is below some minimum threshold distance $d_{min}$, then the robot will actuate and move itself away from $\partial V_i$.  If a robot's distance to $\partial V_i$ is above $d_{min}$, then the robot will execute no control actions. Robots will execute $U_{S_A}$ until they have reached a state where $d(q_{k}, \partial V_i) > d_{min}$ or until they are tasked to leave at a later assignment round.  Similarly, robots will execute $U_{S_P}$ until either $d(q_{k}, \partial V_i) \leq d_{min}$ or they are tasked to leave.  The hybrid robot control policy is given by
\begin{subequations}\label{eq:control}
\begin{align}
U_L(\mathbf{q}_k) & = \bm{\omega}_i \times c\frac{F(\mathbf{q}_k)}{\Vert F(\mathbf{q}_k)\Vert}, \\
U_{S_A}(\mathbf{q}_k) & = -\bm{\omega}_i \times c\frac{F(\mathbf{q}_k)}{\Vert F(\mathbf{q}_k)\Vert}, \\
U_{S_P}(\mathbf{q}_k) & = 0.
\end{align}
\end{subequations}
Here, $\mathbf{\omega}_i = [0, \, 0, \, 1 ]^T$ denotes counterclockwise rotation with respect to the centroid of $V_i$, with clockwise rotation being denoted by the negative and $c$ is a constant that sets the linear speed of the robots.  The hybrid control policy generates a control input perpendicular to the velocity of the fluid as measured by robot $k$\footnote{The inertial velocity of the fluid can be computed from the robot's flow-relative velocity and position.} and pushes the robot towards $\partial V_i$ if $U_L$ is selected, away from $\partial V_i$ if $U_{S_A}$ is selected, or results in no control input if $U_{S_P}$ is selected  The hybrid control policy is summarized by Algorithm \ref{alg:auction} and Fig. \ref{fig:controller}.

\begin{algorithm}                   
\begin{centering}
\caption{Auction Phase \label{alg:auction}}          
\begin{algorithmic}[1]
  \IF {$ElapsedTime == T_a$}
      \STATE Determine $N_i(t)$ and $Q_i$
      \STATE {$\forall k \in Q_i$} 
          \IF {$N_i(t) > \bar{N}_i$}
              \IF {$k \in Q_L$}
                  \STATE {$u_k \gets U_L$}
              \ELSE
              \STATE {$u_k \gets U_S$}
              \ENDIF
          \ELSE
              \STATE {$u_k \gets U_S$}
          \ENDIF
  \ENDIF
\end{algorithmic}
\end{centering}
\end{algorithm}

In general, the Auction Phase is executed at a frequency of $1/T_a$ which means robots also switch between controller states at a frequency of $1/T_a$.  To further reduce actuation efforts exerted by each robot, it is possible to limit a robot's actuation time to a period of time $T_c \leq T_a$. Such a scheme may prolong the amount of time required for the team to achieve the desired allocation, but may result in significant energy-efficiency gains.  We further analyze the proposed strategy in the following sections.

\section{Analysis}\label{sec:analysis}
In this section, we discuss the theoretical feasibility of the proposed
distributed allocation strategy.  Instead of the traditional agent-based
analysis, we employ a macroscopic analysis of the proposed distributed control
strategy given by Algorithm \ref{alg:auction} and \eqref{eq:control}.  We
  first note that while the single robot controller shown in
  Fig. \ref{fig:controller} results in an agent-level stochastic control
  policy, the ensemble dynamics of a team of $N$ robots each executing the
  same hybrid control strategy can be modeled using  a {\it polynomial
  stochastic hybrid system} (pSHS). The advantage of this approach is that it
allows the use of moment closure techniques to model the time evolution of the
distribution of the team across the various cells. This, in turn, enables
  the analysis of the stability of the closed-loop ensemble dynamics.  The
  technique was previously illustrated in \citep{ref:Mather11}.  For completeness, we briefly summarize the approach here and refer the interested reader to \citep{ref:Mather11} for further details.

The system state is given by $\mathbf{N}(t) = \left[N_1(t), \ldots,
  N_M(t)\right]^T$.  As the team distributes across the $M$ regions, the rate
in which robots leave a given $V_i$ can be modeled using constant {\it
  transition rates}.  For every edge $e_{ij} \in {\cal E}$, we assign a
constant $a_{ij} > 0$ such that $a_{ij}$ gives the transition probability per
unit time for a robot from $V_i$ to land in $V_j$.  Different from
\cite{ref:Mather11}, the $a_{ij}$s are a function of the parameters $c$,
$T_c$, and $T_a$ of the individual robot control policy \eqref{eq:control},
the dynamics of the surrounding fluid, and the inherent noise in the
environment.  Furthermore, $a_{ij}$ is a macroscopic description of the system
and thus a parameter of the ensemble dynamics rather than the agent-based
system.  As such, the macroscopic analysis is a description of the
steady-state behavior of the system and becomes exact as $N$ approaches infinity.

Given ${\cal G}$ and the set of $a_{ij}$s, we model the ensemble dynamics as a set of transition rules of the form:
\begin{eqnarray}\label{eq:rxns}
N_i \xrightarrow{a_{ij}} N_j \quad \textrm{$\forall$ $e_{ij} \in {\cal E}$}.
\end{eqnarray}
The above expression represents a stochastic transition rule with $a_{ij}$ as the per unit transition rate and $N_i(t)$ and $N_j(t)$ as discrete random variables.  In the robotics setting, \eqref{eq:rxns} implies that robots at $V_i$ will move to $V_j$ with a rate of $a_{ij}N_i$.  We assume the ensemble dynamics is Markovian and note that in general $a_{ij} \neq a_{ji}$ and $a_{ij}$ encodes the inverse of the average time a robot spends in $V_i$.

Given \eqref{eq:rxns} and employing the extended generator we can obtain the following description of the moment dynamics of the system:
\begin{align}\label{eq:ensembleDyn}
\tfrac{d}{dt}\E[\mathbf{N}] &= \A \E[\mathbf{N}]
\end{align}
where $[\A]_{ij} =  a_{ji}$ and $[\A]_{ii} = -\sum_{(i,j)\in\cal{E}} a_{ij}$ \citep{ref:Mather11}. It is important to note that $\A$ is a Markov process matrix and thus is negative semidefinite.  This, coupled with the conservation constraint $\sum_i N_i = N$ leads to exponential stability of the system given by \eqref{eq:ensembleDyn} \citep{ref:Klavins10}.

In this work, we note that $a_{ij}$s can be determined experimentally after the selection of the various parameters in the distributed control strategy.  While the $a_{ij}$s can be chosen to enable the team of robots to autonomously maintain the desired steady-state distribution \citep{ref:Hsieh08b}, extraction of the control parameters from user specified transition rates is a direction for future work.  Thus, using the technique described by \citet{ref:Mather11}, the following result can be stated for our current distributed control strategy

\begin{theorem}\label{thm:1} Given a team of $N$ robots with kinematics given by \eqref{eq:kinematics} and $\mathbf{v}_f$ given by \eqref{eq:doubleGyre}, the distributed allocation strategy given by Algorithm \ref{alg:auction} and \eqref{eq:control}, at the ensemble level is stable and achieves the desired allocation strategy.
\end{theorem}
For the details of the model development and the proof, we refer the interested reader to \citep{ref:Mather11}.

\section{Simulation Results}\label{sec:results}
We validate the proposed control strategy described by Algorithm \eqref{alg:auction} and \eqref{eq:control} using three different flow fields:
\begin{enumerate}
\item the time invariant wind driven multi-gyre model given by \eqref{eq:doubleGyre} with $\varepsilon = 0$,
\item the time varying wind driven multi-gyre model given by \eqref{eq:doubleGyre} for a range of $\omega \neq 0$ and $\varepsilon \neq 0$ values, and
\item an experimentally generated flow field using different values of $T_a$ and $c$ in \eqref{eq:control}.
\end{enumerate}
We refer to each of these as Cases 1, 2, and 3 respectively.  Two metrics are used to compare the three cases. The first is the mean vehicle control effort to indicate the energy expenditure of each robot. The second is the population root mean square error (RMSE) of the resulting robot population distribution with respect to the desired population.  The RMSE is used to show effectiveness of the control policy in achieving the desired distribution.

\begin{figure}
\centering
\subfigure{
\includegraphics[width=0.25\linewidth]{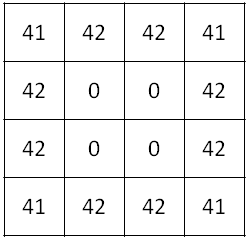}
\label{fig:ringPattern}}
\subfigure{
\includegraphics[width=0.25\linewidth]{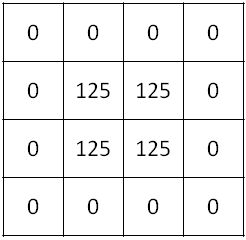}
\label{fig:blockPattern}}
\subfigure{
\includegraphics[width=0.25\linewidth]{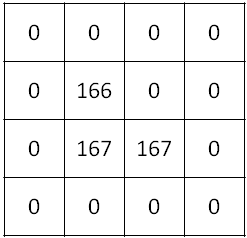}
\label{fig:LPattern}}%
\caption{Three desired distributions of the team of $N=500$ mobile sensing
  resources/robots. (a) A Ring pattern formation, (b) a Block pattern
  formation, and (c) an L-shaped pattern formation. Each box represents a gyre
  and the number designates the desired number of robots contained within each
  gyre.\label{fig:patterns}}
\end{figure}

All cases assume a team of $N = 500$ robots.  The robots are randomly
  distributed across the set of M gyres in ${\cal W}$.  For the theoretical
  models, the workspace ${\cal W}$ consists of a $4 \times 4$ set of gyres,
  and each $V_i \in {\cal V}$ corresponds to a gyre as shown in
  Fig. \ref{fig:gyreTessellation}.  We considered three sets sets of desired
  distributions, namely a Ring formation, a Block formation, and an L-shaped
  formation as shown in Fig. \ref{fig:patterns}.  The experimental flow data
  had a set of $3 \times 4$ regions. The inner two cells comprised ${\cal W}$,
  while the complement, ${\cal W}^C$ consisted of the remaining cells.  This
  designation of cells helped to isolate the system from boundary effects, and
  allowed the robots to escape the center gyres in all directions.  The
  desired pattern for this experimental data set was for all the agents to be contained
  within a single cell.  Each of the three cases was simulated a minimum of five times and for a long enough period of time until steady-state was reached.

\subsection{Case I: Time-Invariant Flows}
\begin{figure}[t]
\centering
\subfigure[No Control]{\includegraphics[width=0.49\linewidth]{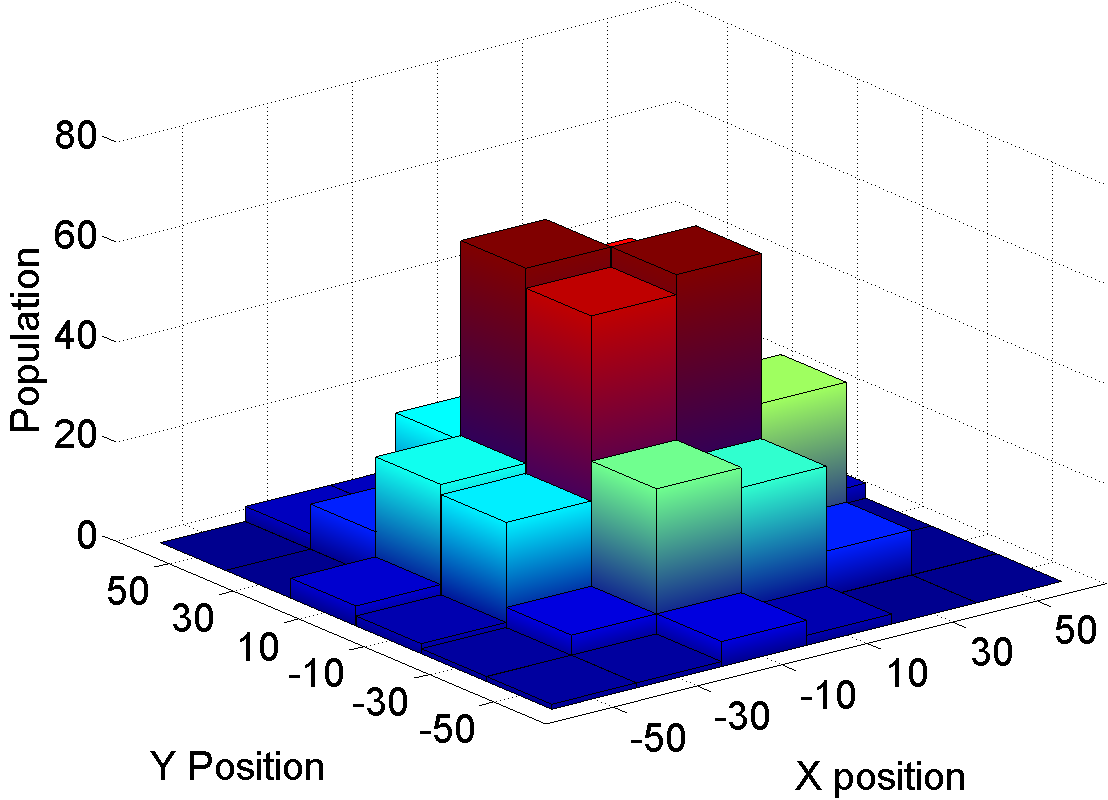}\label{fig:histNoControl}}
\subfigure[Ring]{\includegraphics[width=0.49\linewidth]{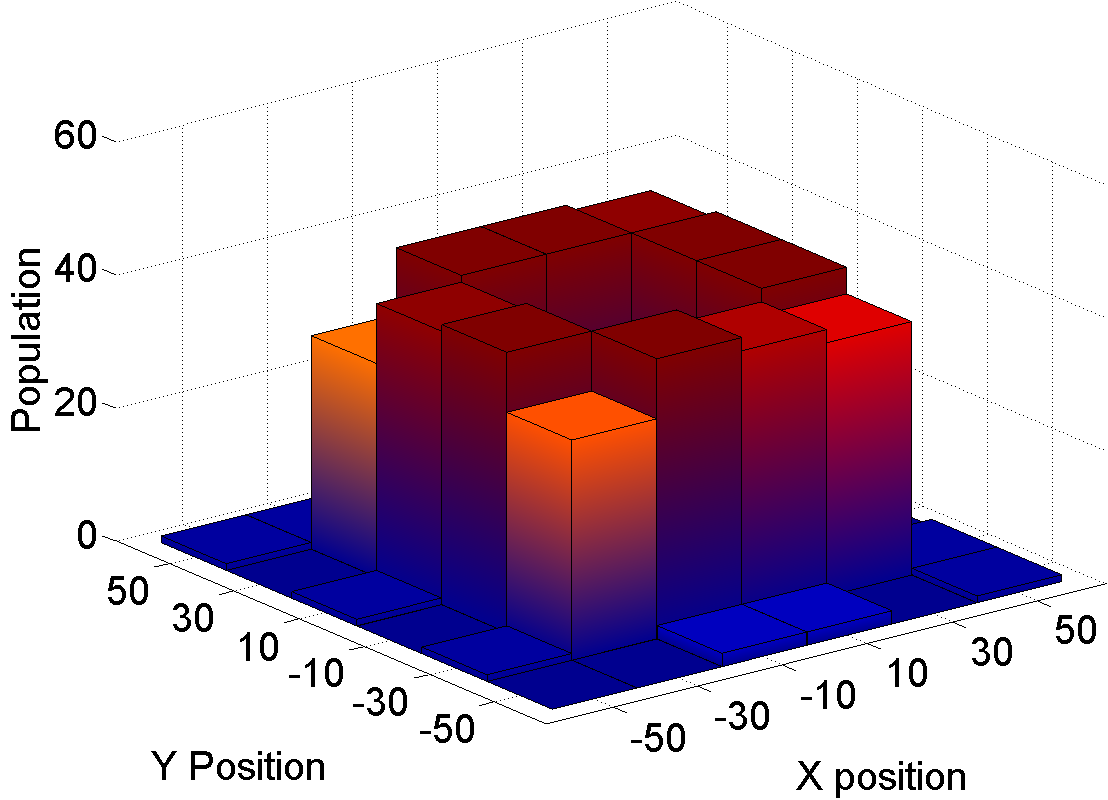}\label{fig:ringPopulation}}
\subfigure[Block]{\includegraphics[width=0.49\linewidth]{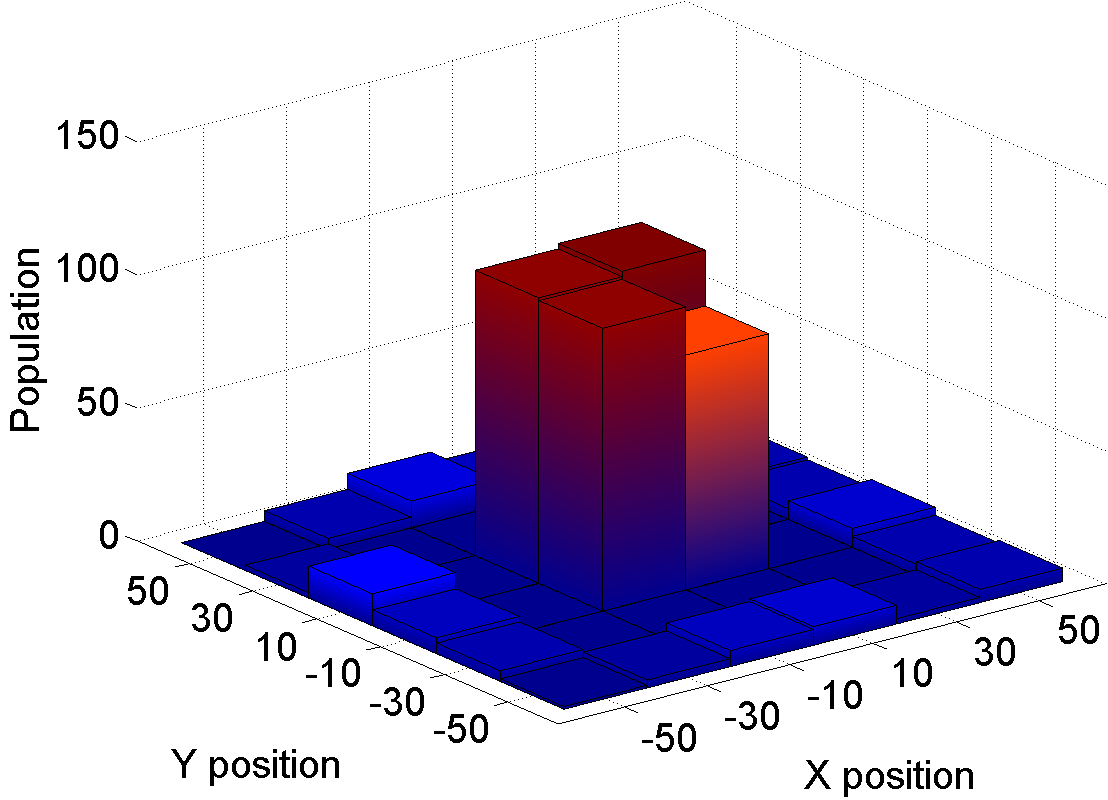}\label{fig:blockPopulation}}
\subfigure[L-Shape]{\includegraphics[width=0.49\linewidth]{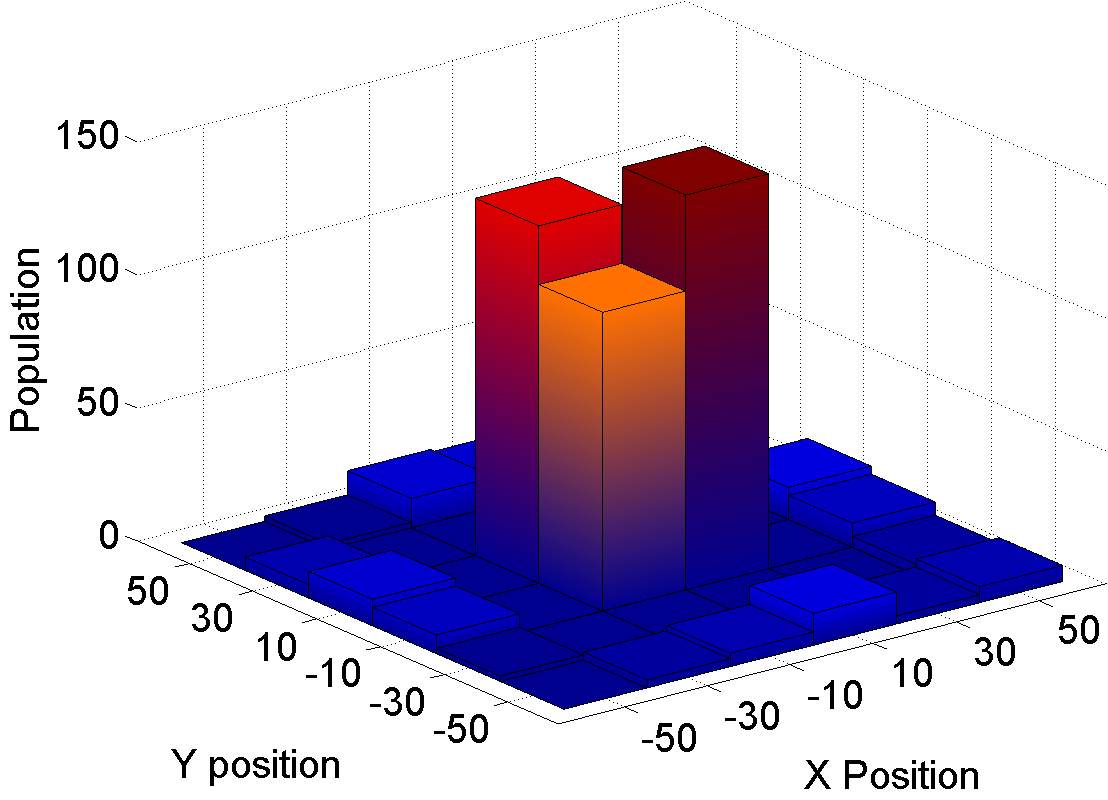}\label{fig:LPopulation}}
\caption{Histogram of the final allocations in the time-invariant flow field for the swarm of (a) passive robots exerting no controls and robots exerting control forming the (b) Ring pattern with $T_c=0.8T_a$ at $t=450$, (c) Block pattern with $T_c=T_a$ at $t=450$, and (d) L-shape pattern with $T_c=0.5T_a$ at $t=450$. \label{fig:finalPDF}}
\end{figure}

For time-invariant flows, we assume $\varepsilon=0$, $A=0.5$, $s=20$, $\mu =
0.005$, and $I=35$ in \eqref{eq:doubleGyre}.  For the ring pattern, we
  consider the case when the actuation was applied for $T_c = f T_a$ amount of
  time where $f = 0.1, 0.2, \ldots, 1.0$, and $T_a=10$.  For the Block and L-Shape patterns, we considered the cases when $T_c = 0.5T_a$ and $T_c = T_a$.  The final population distribution of the team for the case with no controls and the cases with controls for each of the patterns are shown in Fig. \ref{fig:finalPDF}.



\begin{figure}
\centering
\subfigure{\includegraphics[width=0.8\linewidth]{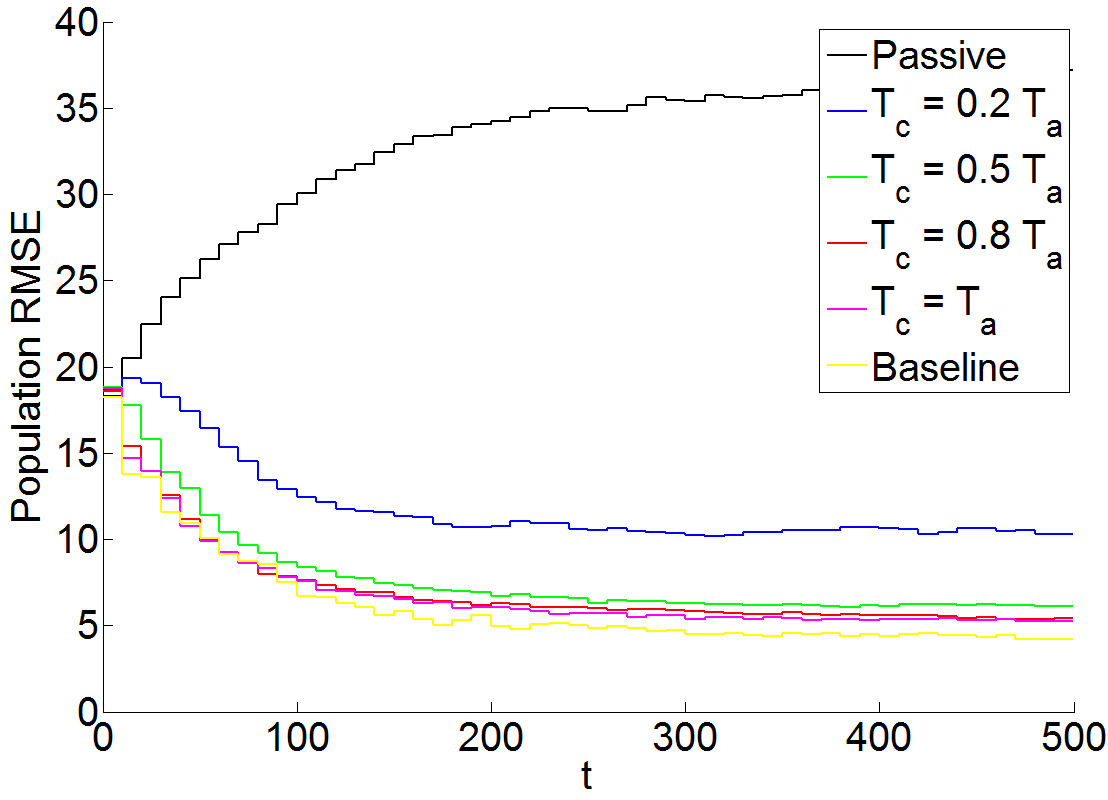}\label{fig:ringRMSE}}
\subfigure{\includegraphics[width=0.8\linewidth]{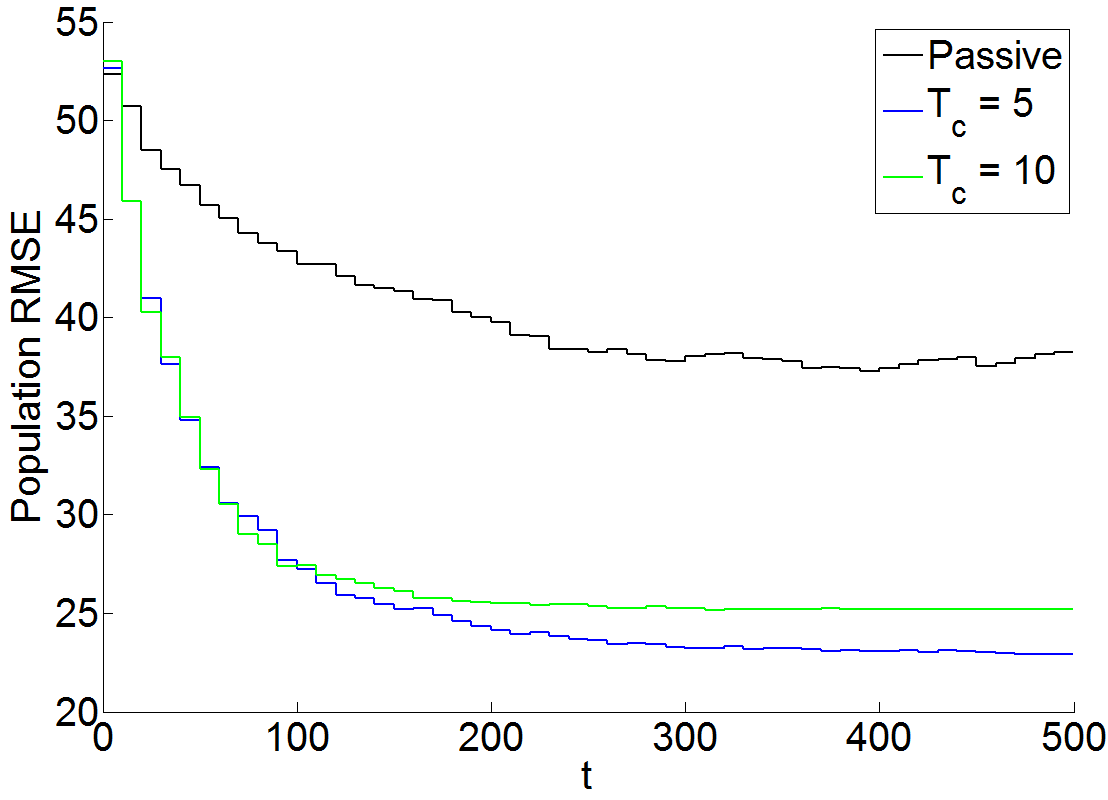}\label{fig:blockRMSE}}
\subfigure{\includegraphics[width=0.8\linewidth]{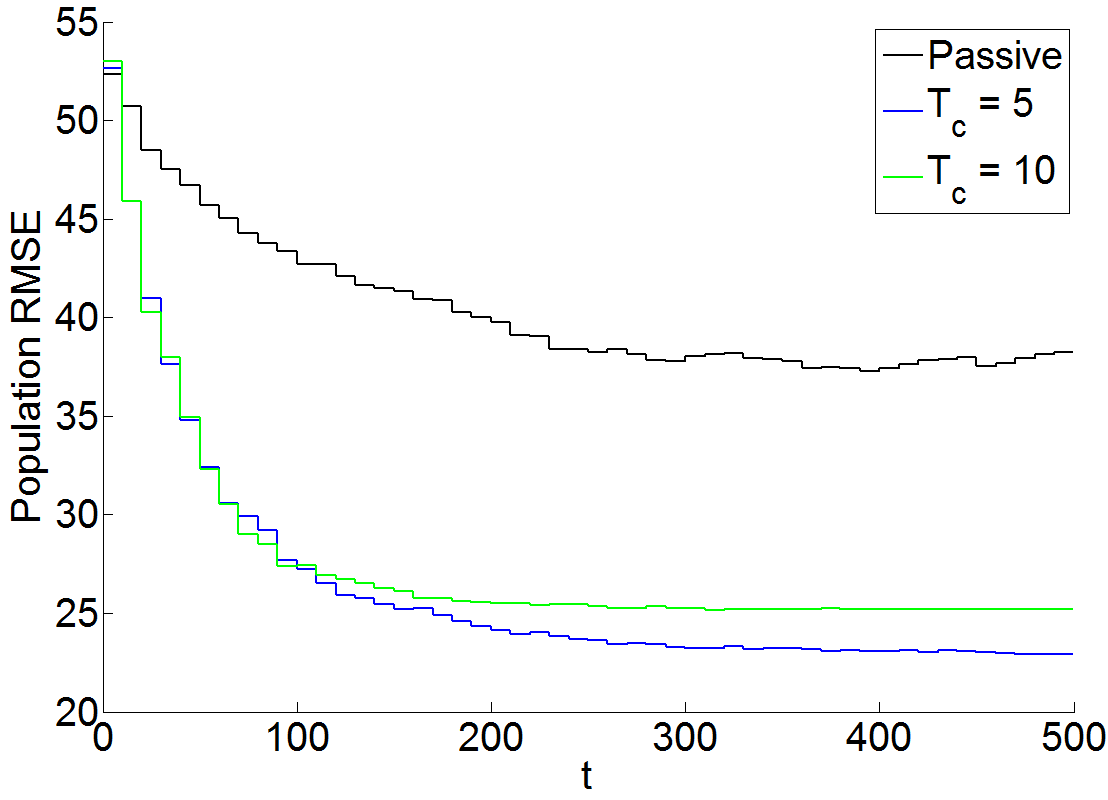}\label{fig:LRMSE}}
\caption{Comparison of the population RMSE in the time-varying flow
  for the (a) Ring formation, (b) the Block formation, and (c) the L-shape
    formation for different $T_c$, and for the PID control baseline controller in the Ring case with time-invariant flows. \label{fig:rmseResults}}
\end{figure}

\begin{figure}
\centering
\includegraphics[width=0.8\linewidth]{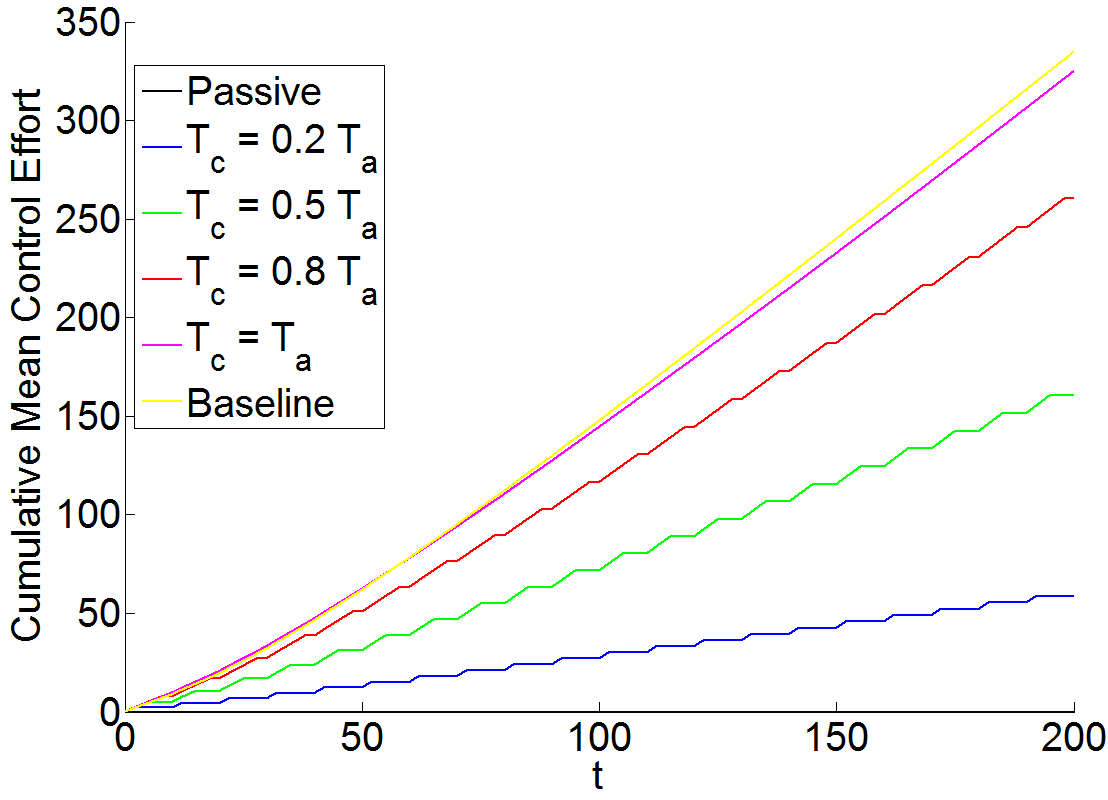}
\caption{Comparison of the total control effort for the {Ring} pattern for different $T_c$ with the baseline controller for time-invariant flows. \label{fig:effortResults}}
\end{figure}

 We compared our results to a baseline deterministic
allocation strategy where the desired allocation is pre-computed and
individual robots follow fixed trajectories when navigating from one gyre to
another.  For this baseline case, robots travel in straight lines at fixed
speeds using a simple PID trajectory follower and treat the surrounding fluid
dynamics as an external disturbance source.  The RMSE results for all patterns
are summarized in Table \ref{RMSEtable} and
Fig.~\ref{fig:rmseResults}. The cumulative control effort per agent is shown in Fig. \ref{fig:effortResults}.  From Fig. \ref{fig:rmseResults}, we see that our proposed control strategy performs comparable to the baseline case especially when $T_c=T_a=10$ $sec$.  In fact, even when $T_c < T_a$, our proposed strategy achieves the desired distribution.  The advantage of the proposed approach lies in the significant energy gains when compared to the baseline case, especially when $T_c < T_a$, as seen in Fig. \ref{fig:effortResults}.  We omit the cumulative control effort plots for the other cases since they are similar to Fig. \ref{fig:effortResults}.


In time-invariant flows, we note that for large enough $T_c$, our proposed distributed control strategy performs comparable to the baseline controller both in terms of steady-state error and convergence time.  As $T_c$ decreases, less and less control effort is exerted and thus it becomes more and more difficult for the team to achieve the desired allocation.  This is confirmed by both the RMSE results summarized in Table \ref{RMSEtable} and Fig.~\ref{fig:ringRMSE}-\ref{fig:LRMSE}.  Furthermore, while the proposed control strategy does not beat the baseline strategy as seen in Fig. \ref{fig:ringRMSE}, it does come extremely close to matching the baseline strategy performance.  while requiring much less control effort as shown in Fig. \ref{fig:effortResults} even at high duty cycles, {\it i.e.}, when $T_c/T_a > 0.5$.


More interestingly, we note that executing the proposed control strategy at 100\% duty cycle, {\it i.e.}, when $T_c = T_a$, in time-invariant flows did not always result in better performance.  This is true for the cases when $T_c=0.5T_a=5$ for the Block and L-shaped patterns shown in Fig. \ref{fig:blockRMSE} and \ref{fig:LRMSE}.  In these cases, less control effort yielded improved performance.  However, further studies are required to determine the critical value of $T_c$ when less control yields better overall performance. In time-invariant flows, our proposed controller can more accurately match the desired pattern while using approximately $20\%$ less effort when compared to the baseline controller.

\subsection{Case II: Time-Varying Flows}
\begin{figure}
\centering
\subfigure[No Control]{\includegraphics[width=0.49\linewidth]{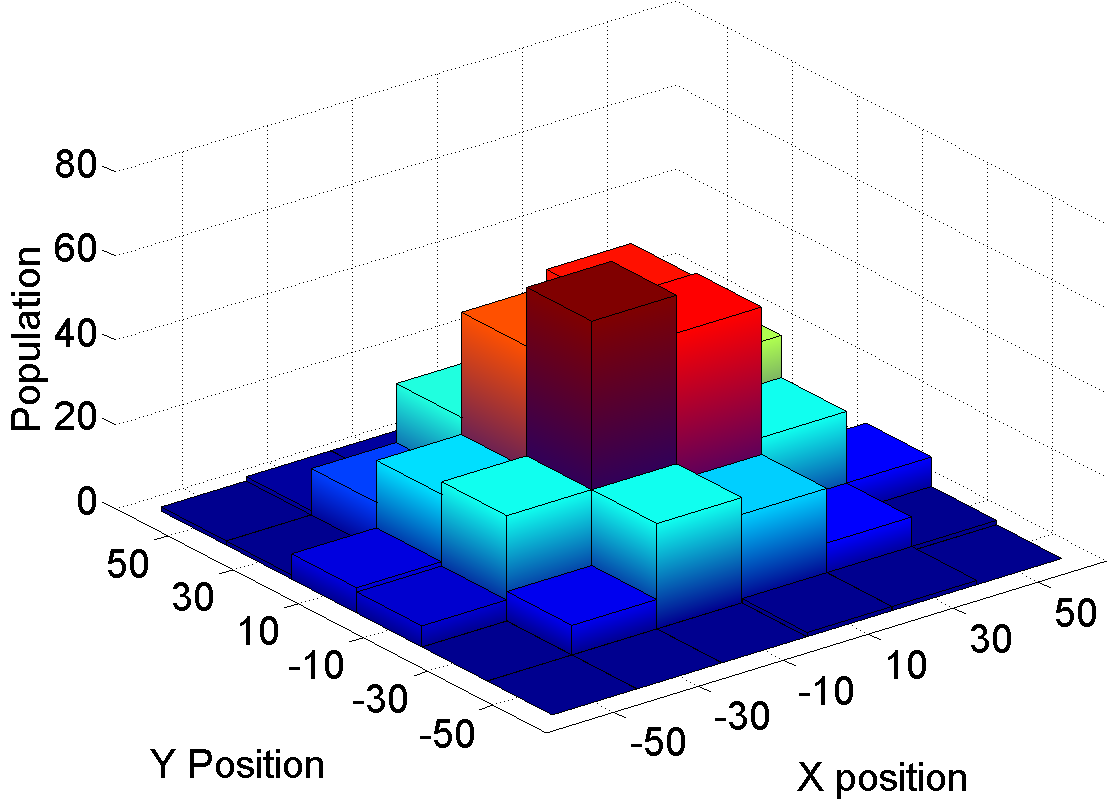}\label{fig:histNoControlTimeVar}}
\subfigure[Ring]{\includegraphics[width=0.49\linewidth]{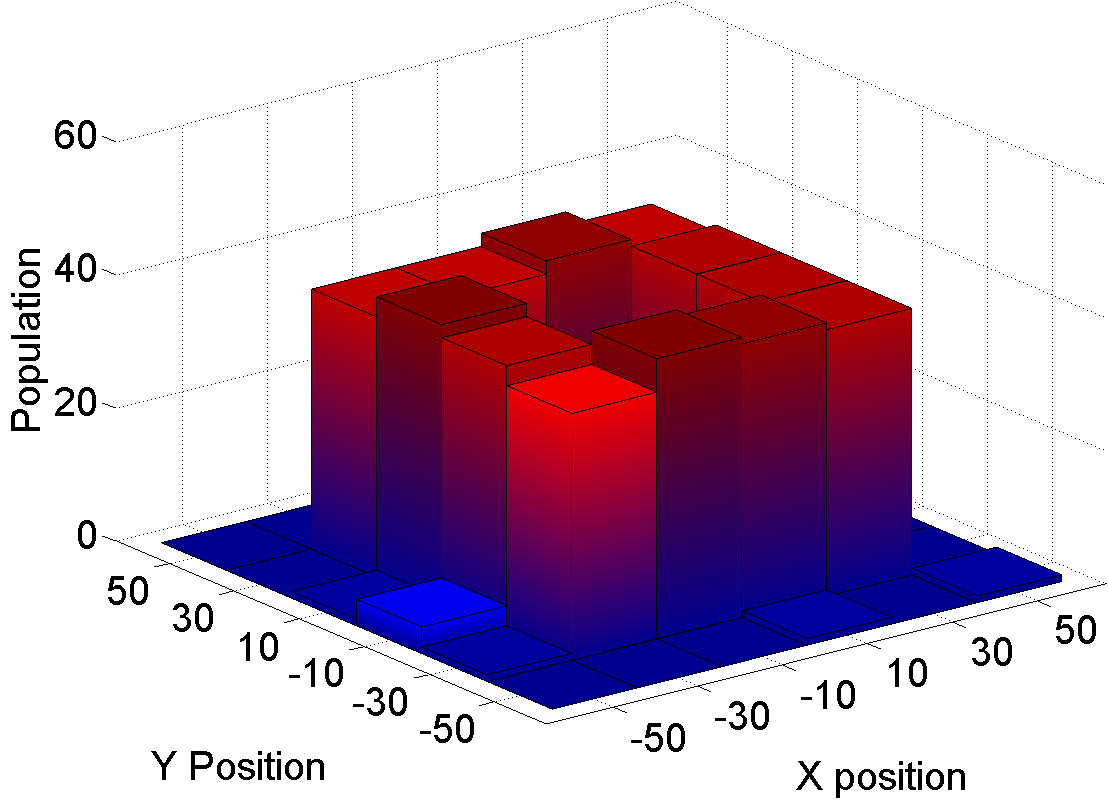}\label{fig:ringPopTimeVar}}
\subfigure[L-Shape]{\includegraphics[width=0.49\linewidth]{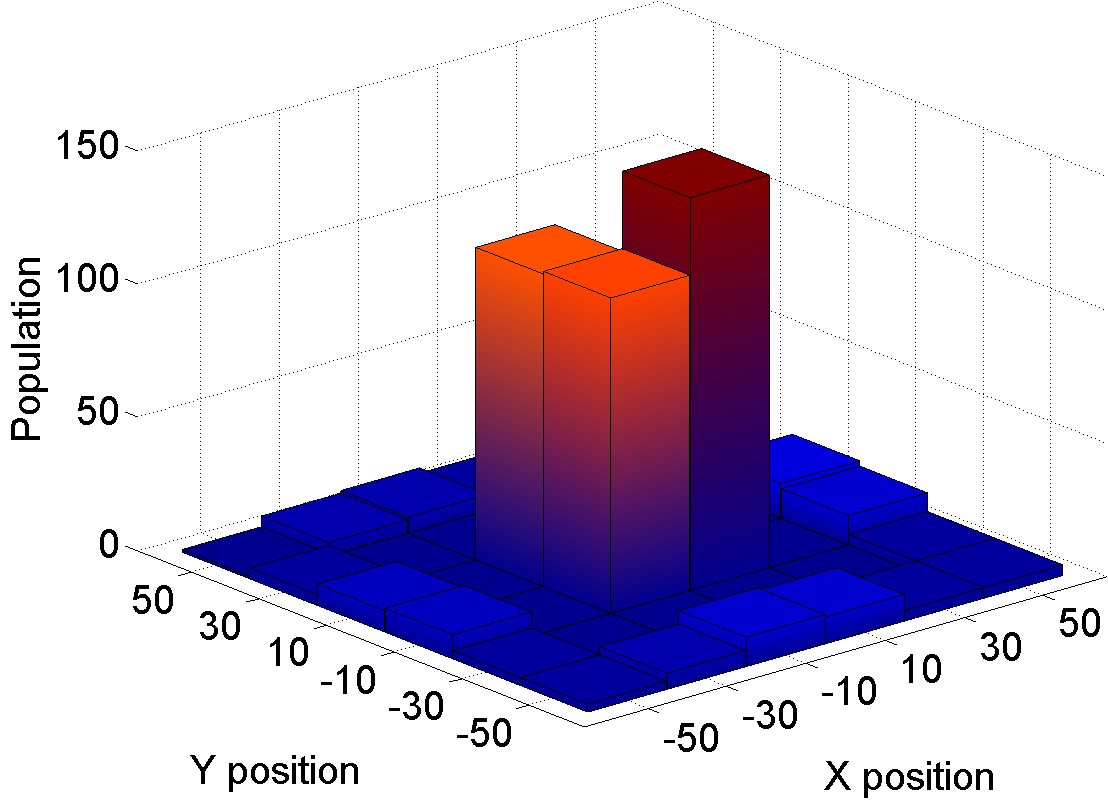}\label{fig:LPopTimeVar}}
\caption{Histogram of the final allocations in periodic flows, with parameters of $\omega = \frac{5*\pi}{40}$ and $\varepsilon = 5$, for the swarm of (a) passive robots exerting no controls and robots exerting control forming the (b) Ring pattern with $T_c=0.8T_a$ at $t=450$, and (c) L-Shape pattern with $T_c=0.5T_a$ at $t=450$. \label{fig:finalPDFVarying}}
\end{figure}

\begin{figure}
\centering
\subfigure[Ring]{\includegraphics[width=0.8\linewidth]{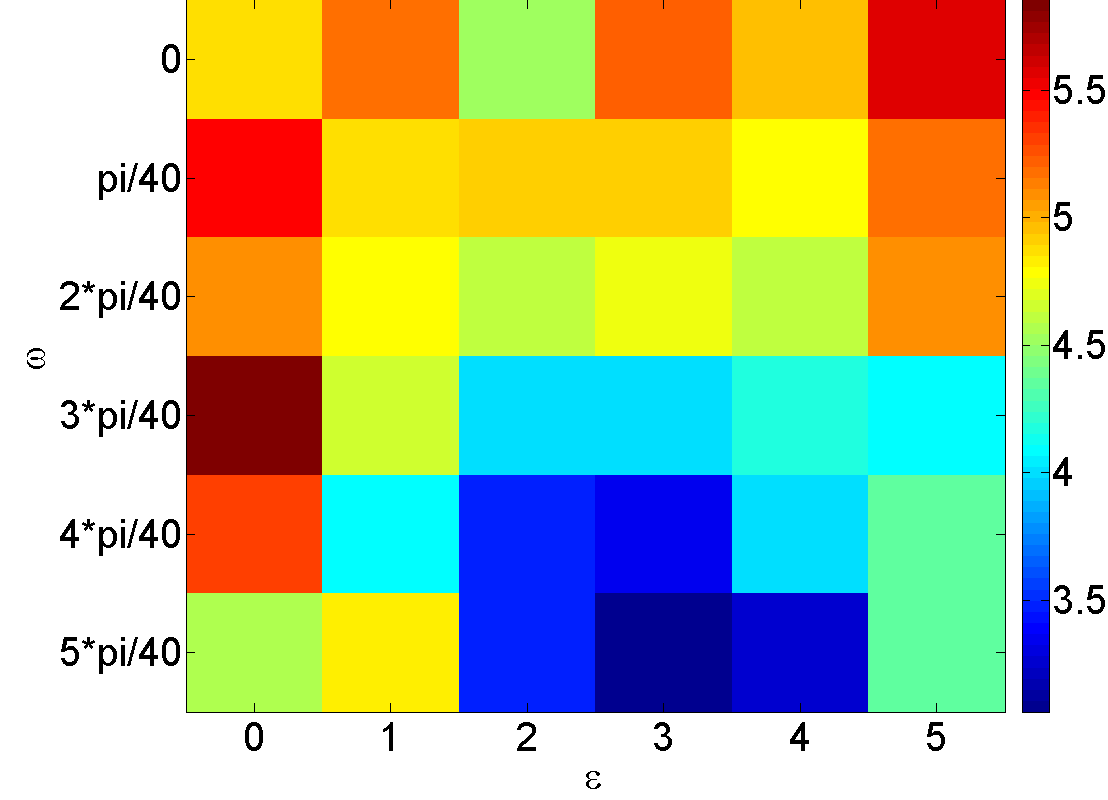}}
\subfigure[L-Shape]{\includegraphics[width=0.8\linewidth]{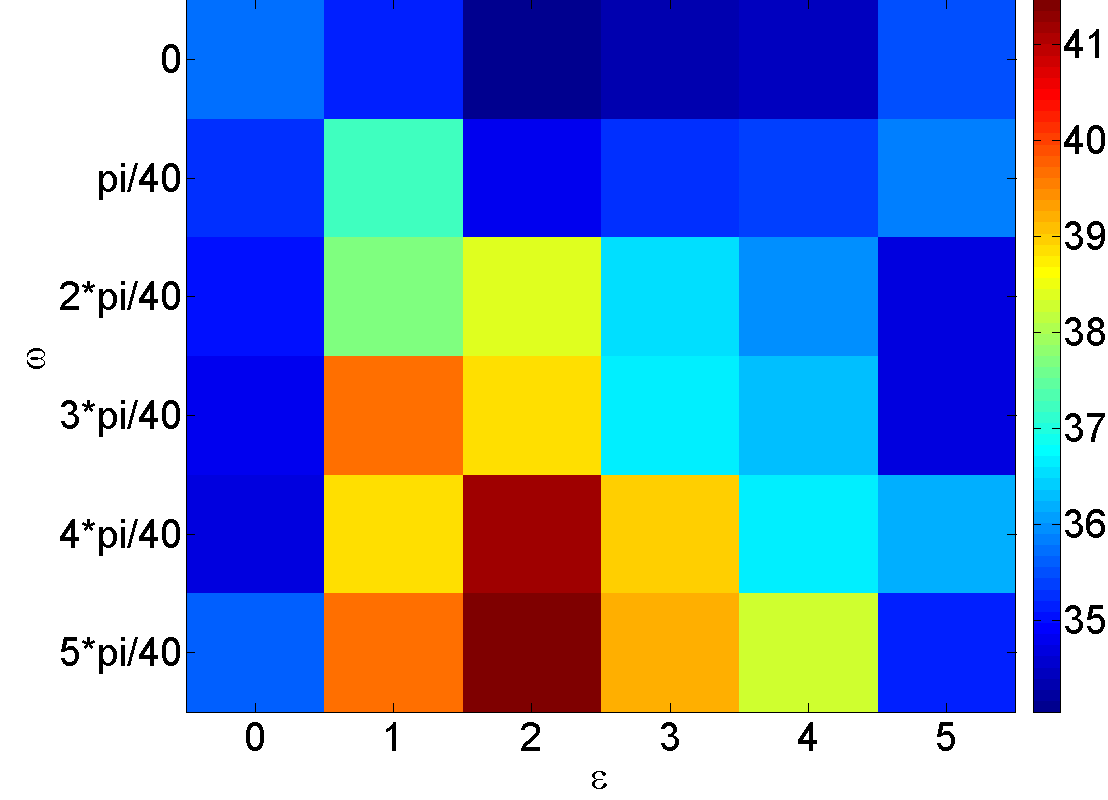}}
\caption{Final population RMSE for different values of $\omega$ and
  $\varepsilon$ for (a) the Ring formation and (b) the L-shaped formation. \label{fig:minVarRMSE}}
\end{figure}

\begin{figure}
\centering
\subfigure{\includegraphics[width=0.8\linewidth]{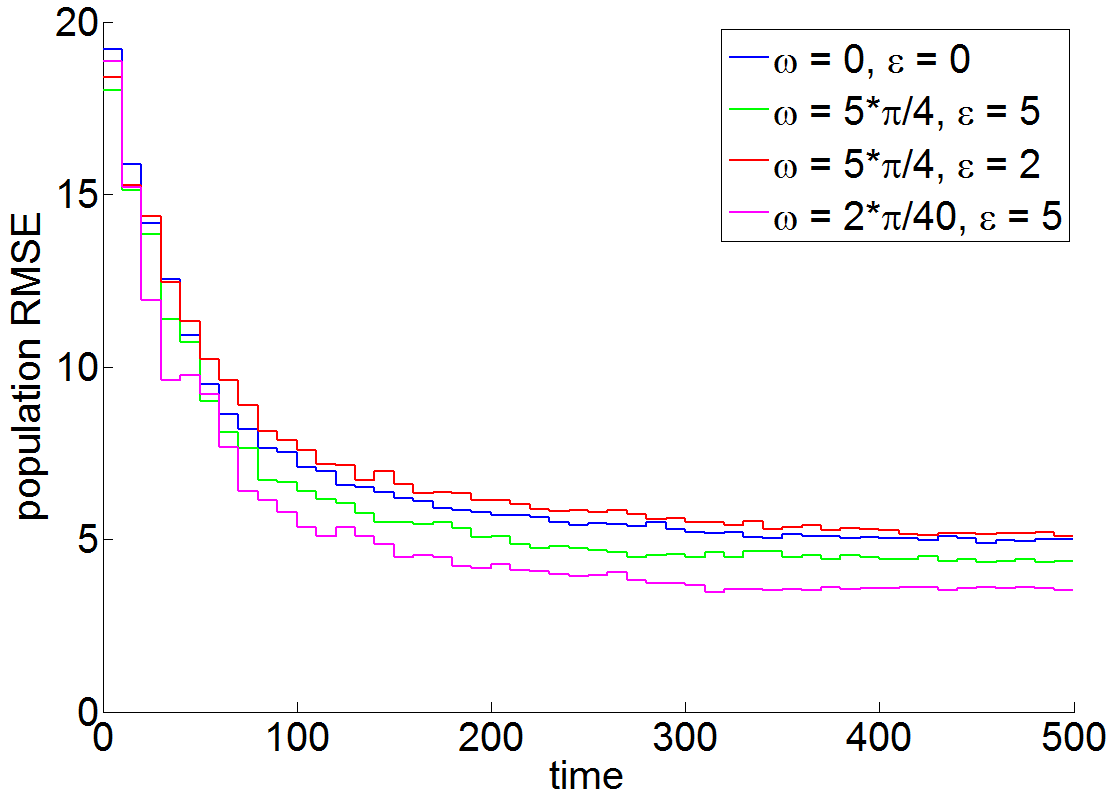}}
\subfigure{\includegraphics[width=0.8\linewidth]{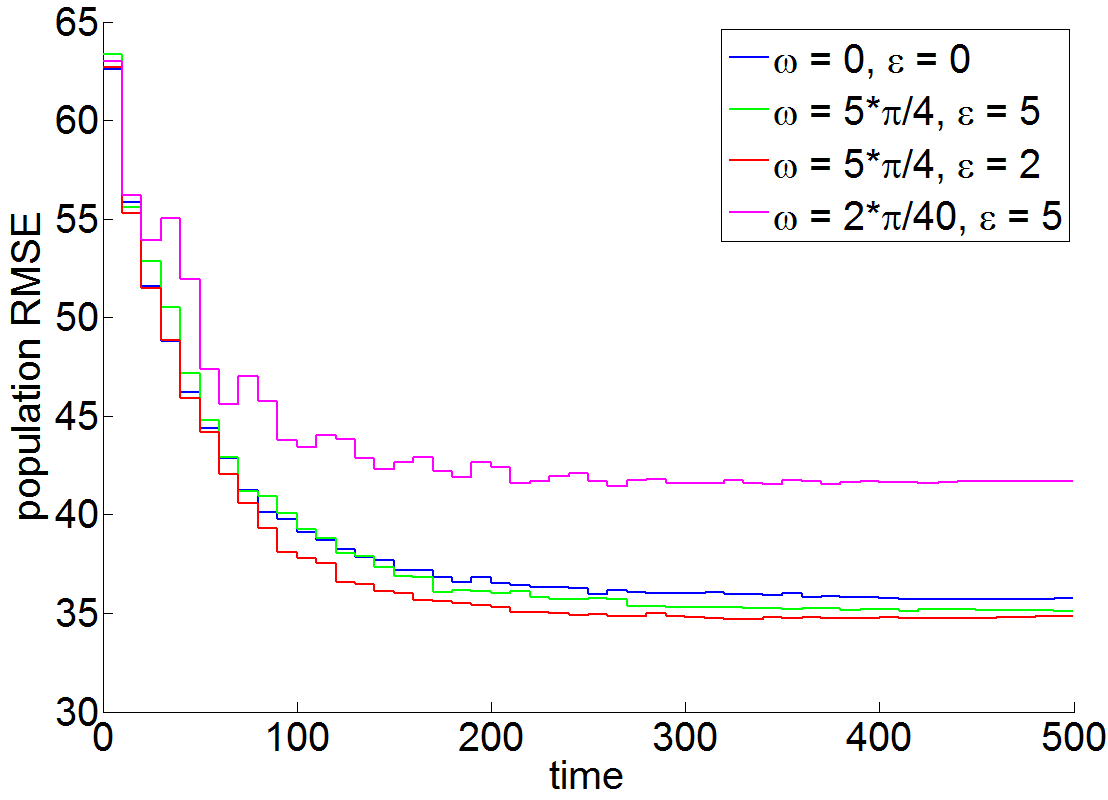}}
\caption{Comparison of RMSE over time for select $\omega$ and $\varepsilon$
  pairs for the (a) Ring and (b) L-shaped pattern{s} in periodic flows. \label{fig:timeVarResultsRMSE}}
\end{figure}

For the time-varying, periodic flow, we assume $A=0.5$, $s=20$, $\mu = 0.005$,
$I=35$, and $\psi = 0$ in
\eqref{eq:doubleGyre}. Additionally, we considered the performance of our
control strategy for different values of $\omega$ and $\varepsilon$ with
$T_a=10$ and $T_c
= 8$ for the Ring formation and $T_c=5$ for the L-shaped formation.  In all these simulations, we use the FTLE ridges obtained for the time-independent case to define the boundaries of each $V_i$.  The final population
distribution of the team for the case with no controls and the cases with
controls for the Ring and L-shape patterns are shown in Fig. \ref{fig:finalPDFVarying}.  The final population RMSE for the cases with different $\omega$ and $\varepsilon$ values for the Ring and L-shape patterns are shown in Fig. \ref{fig:minVarRMSE}.  These figures show the average of $10$ runs for each $\omega$ and $\varepsilon$ pair.  In each of these runs, the swarm of mobile sensors were initially randomly distributed within the grid of $4 \times 4$ cells. Finally, Fig. \ref{fig:timeVarResultsRMSE}  shows the population RMSE as a function of time for the Ring and L-shape patterns respectively.

In time-varying, periodic flows we note that our proposed control strategy is able to achieve the desired final allocation even at 80\% duty cycle, {\it i.e.}, $T_c=0.8T_a$.  This is supported by the results shown in Fig. \ref{fig:minVarRMSE}.  In particular, we note that the proposed control strategy performs quite well for a range of $\omega$ and $\varepsilon$ parameters for both the Ring and L-shape patterns. While the variation in final RMSE values for the Ring pattern is significantly lower than the L-shape pattern, the variations in final RMSE values for the L-shape are all within 10\% of the total swarm population.

\subsection{Case III: Experimental Flows}
\begin{figure}
\centering
\subfigure[]{\label{fig:pivTank}\includegraphics[width=0.67\linewidth]{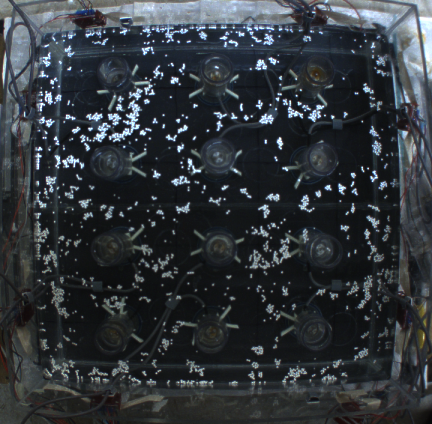}}\\
\subfigure[]{\label{fig:pivResult}\includegraphics[width=0.70\linewidth]{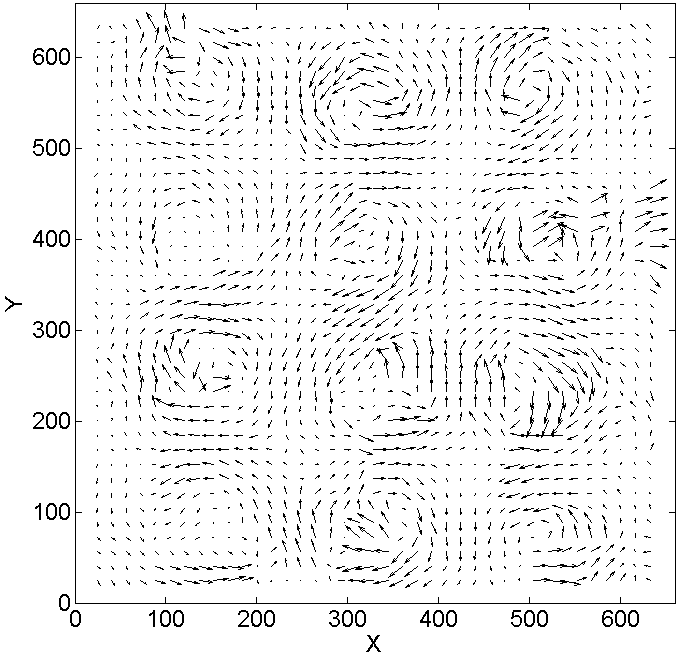}}
\caption{(a) Experimental setup of flow tank with $12$ driven cylinders. (b) Flow field for image (a) obtained via particle image velocimetry (PIV). \label{fig:expSetup}}
\vspace{-10pt}
\end{figure}

Using our $0.6 m \times 0.6 m \times 0.3 m$ experimental flow tank equipped
with a grid of $4 x 3$ set of driving cylinders, we generated a time-invariant
multi-gyre flow field to use in simulation.  Particle image velocimetry
(PIV) was used to extract the surface flows at $7.5$ $Hz$ resulting in a
$39 \times 39$ grid of velocity measurements.  The data was collected for a
total of $60$ $sec$.  Figure \ref{fig:expSetup} shows the top view of our
experimental testbed and the resulting flow field obtained via PIV. Further
  details regarding the experimental testbed can be found in \citep{ref:michiniDSC2013}. Using this data, we simulated a swarm of $500$ mobile sensors executing the control strategy given by \eqref{eq:control}.  

\begin{figure}
\centering
\includegraphics[width=0.65\linewidth]{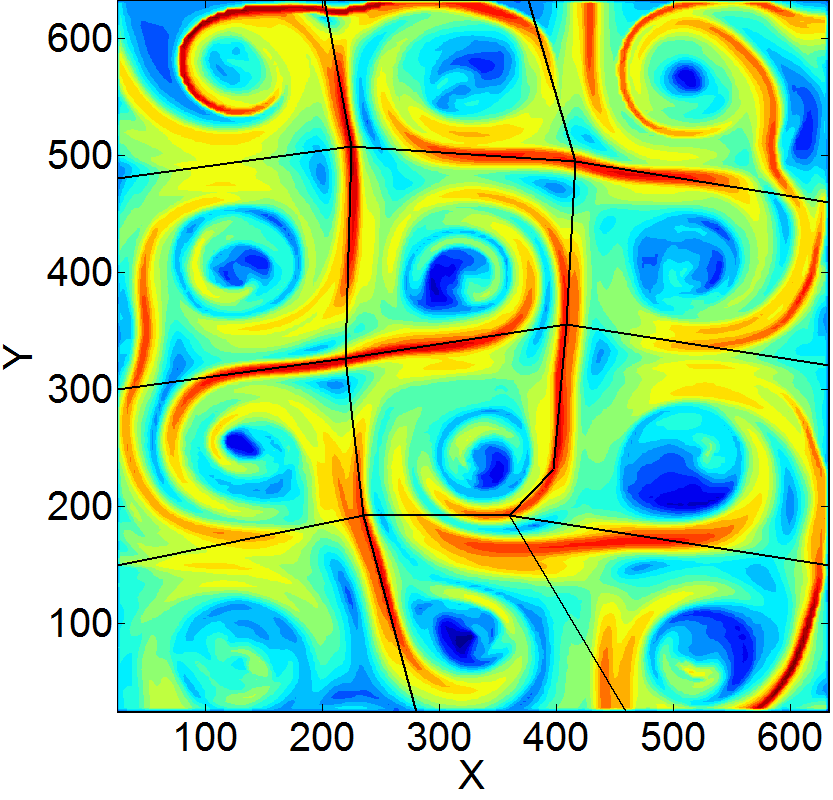}
\caption{FTLE field for the temporal mean of the experimental velocity data.  The field is discretized into a grid of $4 \times 3$ cells whose boundaries are shown in black.\label{fig:tankCells}}
\end{figure}

\begin{figure}
\centering
\subfigure[]{\includegraphics[width=0.45\linewidth]{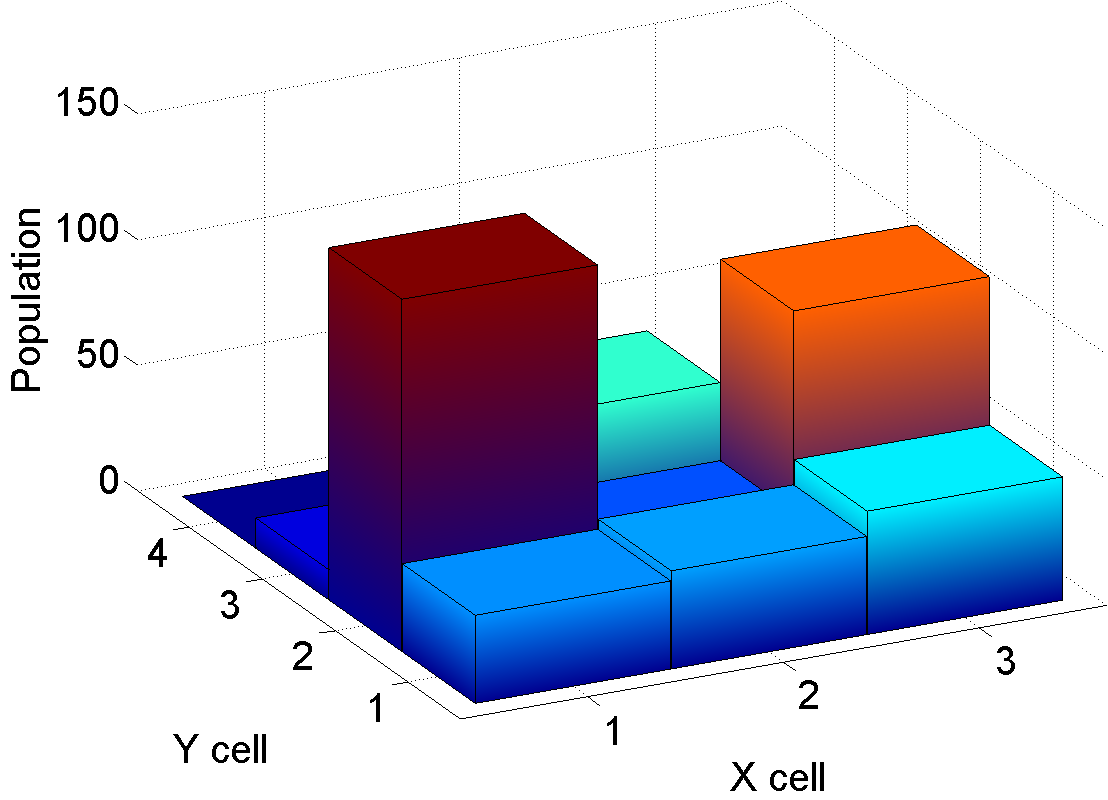}\label{fig:tankPassive}}
\subfigure[]{\includegraphics[width=0.45\linewidth]{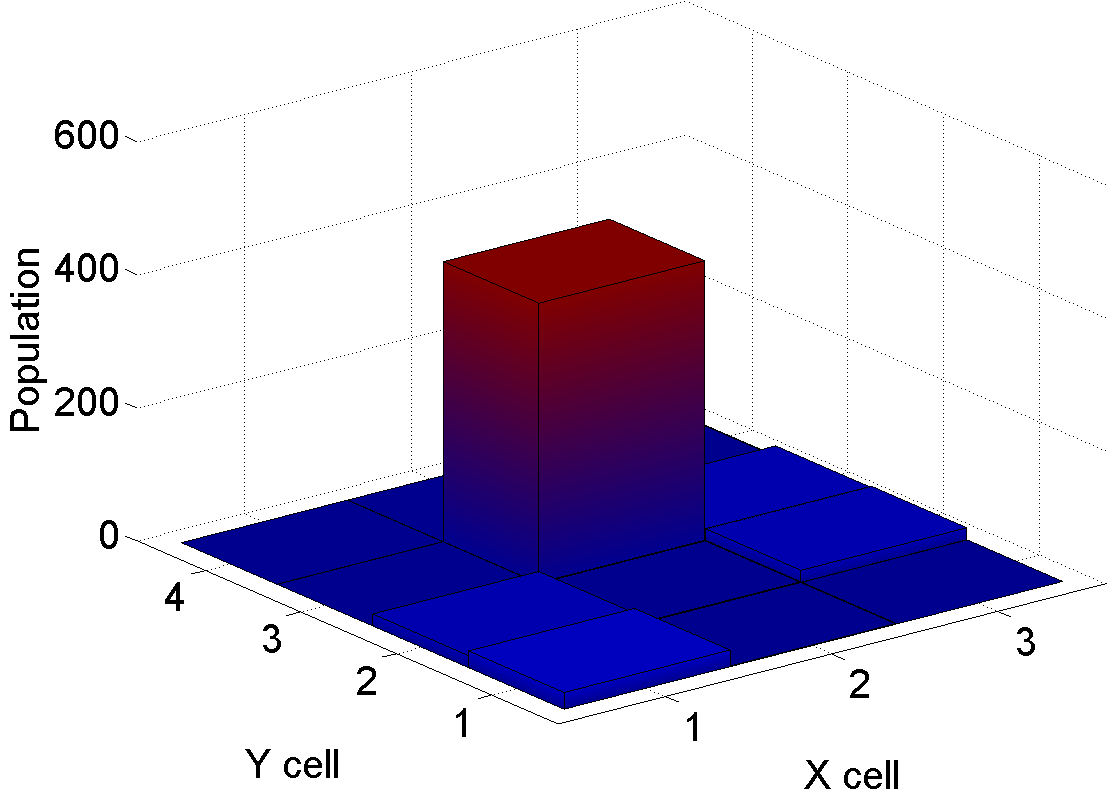}\label{fig:tankPop}}
\caption{Population distribution for a swarm of $500$ mobile sensors over a
  period of $60$ $sec$ (a) with no controls, {\it i.e.}, passive, and (b) with
  controls with $T_c=0.8T_a$.}
\end{figure}

\begin{figure}
\centering
\subfigure[]{\includegraphics[width=0.8\linewidth]{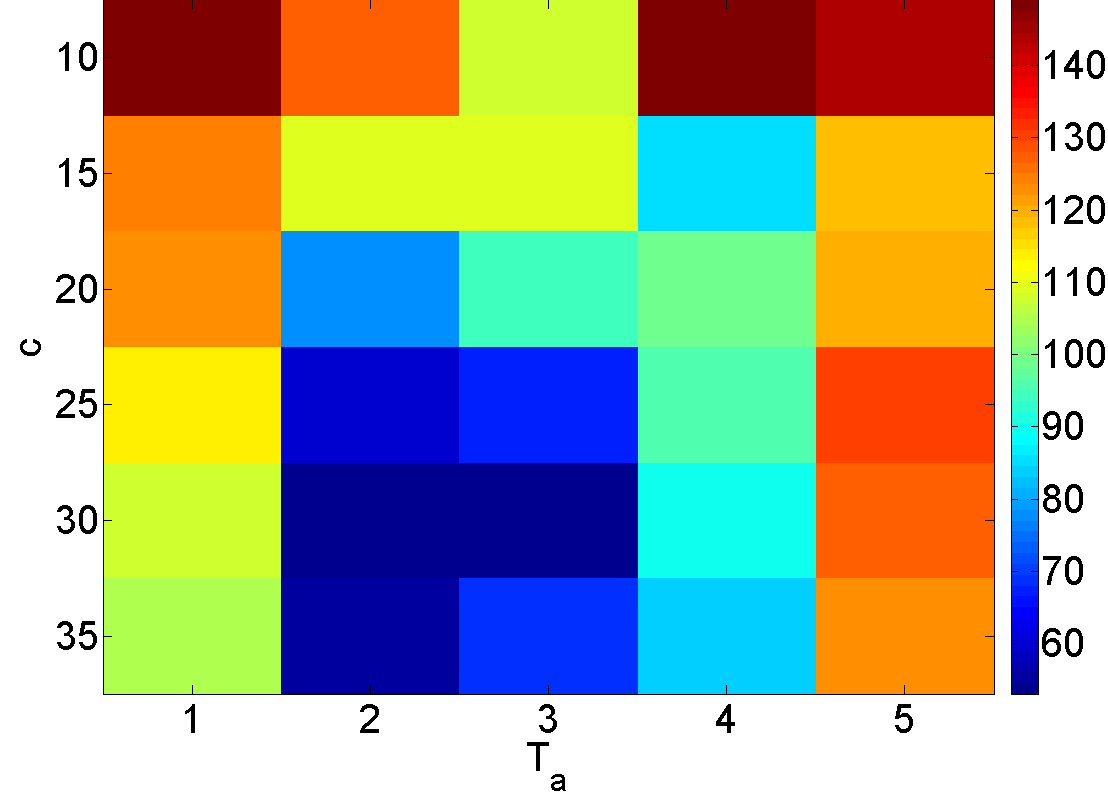}\label{fig:tankFinalRMSE}}
\subfigure[]{\includegraphics[width=0.8\linewidth]{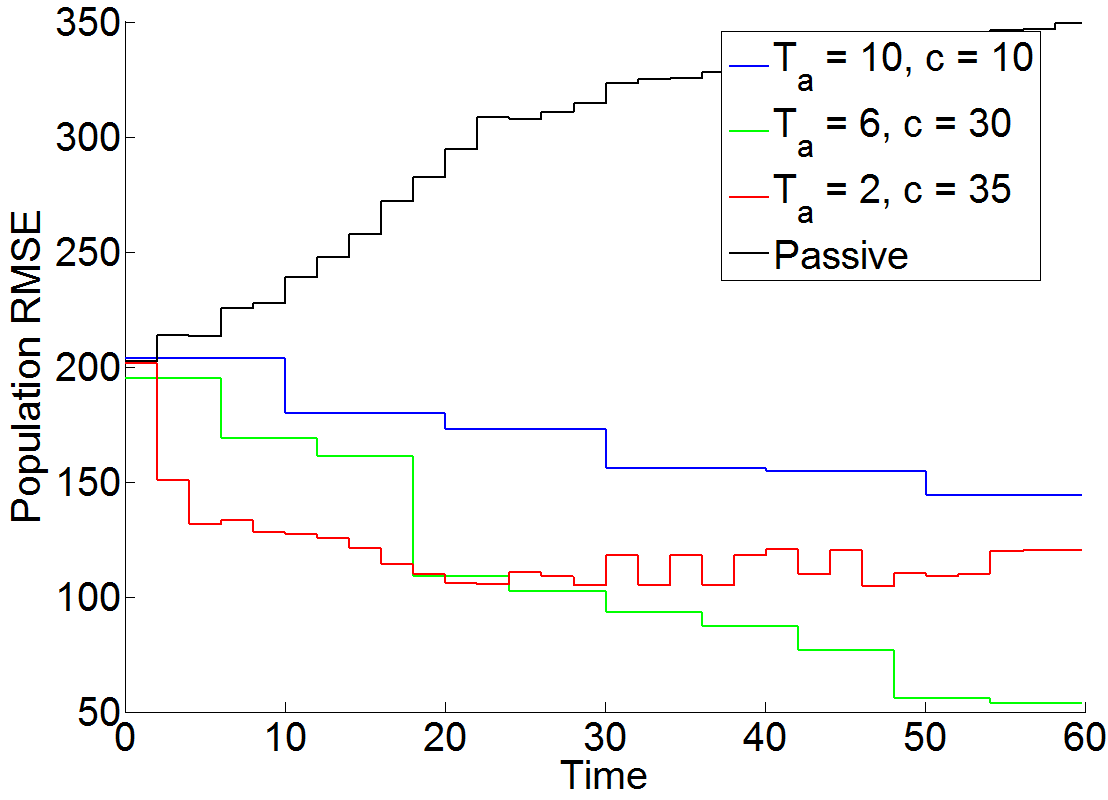}\label{fig:tankTimeRMSE}}
\caption{(a) Final RMSE for different values of $c$ and $T_a$ using the experimental flow field.  $T_c/T_a = 0.8$ is kept constant throughout. (b) RMSE over time for select $c$ and $T_a$ parameters on an experimental flow field. The duty cycle $T_c/T_a = 0.8$ is kept constant throughout.}
\end{figure}

To determine the appropriate tessellation of the workspace, we used the LCS
ridges obtained for the temporal mean of the velocity field. This resulted in the discretization of the space into a grid of $4 \times 3$ cells.  Each cell corresponds to a single gyre as shown Fig. \ref{fig:tankCells}.  The cells of primary concern are the central pair and the remainder boundary cells were not used to avoid boundary effects and to allow robots to escape the center gyres in all directions.  The robots were initially uniformly distributed across the two center cells and all $500$ robots were tasked to stay within the upper center cell.  When no control effort is exerted by the robots, the final population distribution achieved by the team is shown in Fig. \ref{fig:tankPassive}.  With controls, the final population distribution is shown in Fig. \ref{fig:tankPop}.  The control strategy was applied assuming $T_c/T_a = 0.8$. The final RMSE for different values of $c$ in \eqref{eq:control} and $T_a$ is shown in Fig. \ref{fig:tankFinalRMSE} and RMSE as a function of time for different values of $c$ and $T_a$ are shown in Fig. \ref{fig:tankTimeRMSE}.

The results obtained using the experimental flow field shows that the proposed control strategy has the potential to be effective in realistic flows.  However, the resulting performance will require good matching between the amount of control effort a vehicle can realistically exert, the frequency in which the auctions occur within a cell, and the time scales of the environmental dynamics as shown in in Figs. \ref{fig:tankFinalRMSE} and \ref{fig:tankTimeRMSE}.  This is an area for future investigation.

\section{Conclusions and Future Outlook}\label{sec:future}
In this work, we presented the development of a distributed hybrid control strategy for a team of robots to maintain a desired spatial distribution in a stochastic geophysical fluid environment.  We assumed robots have a map of the workspace which in the fluid setting is akin to having some estimate of the global fluid dynamics.  This can be achieved by knowing the locations of the material lines within the flow field that separate regions with distinct dynamics.  Using this knowledge, we leverage the surrounding fluid dynamics and inherent environmental noise to synthesize energy efficient control strategies to achieve a distributed allocation of the team to specific regions in the workspace.  Our initial results show that using such a strategy can yield similar performance as deterministic approaches that do not explicitly account for the impact of the fluid dynamics while reducing the control effort required by the team.

For future work we are interested in using actual ocean flow data to further
evaluate our distributed allocation strategy in the presence of jets and
  eddies~\citep{roger99,miller02,kuznet02,mancho08,branick11,menman2012}.  We
also are interested
in using more complicated flow models including a bounded single-layer PDE
ocean model~\citep{fbys11}, a multi-layer PDE ocean model~\citep{ref:Wang2009,
  ref:Lolla2012}, and realistic 2D and 3D unbounded flow models provided by
the Navy Coastal Ocean Model (NCOM) database.  Particularly, we are interested
in extending our strategy to non-periodic, time-varying flows.  In addition, we are currently developing an experimental testbed capable of generating complex 2D flows in a controlled laboratory setting.  The objective is to be able to evaluate the proposed control strategy using experimentally generated flow field data whose dynamics are similar to realistic ocean flows.  Finally, since our proposed strategy requires robots to have some estimate of the global fluid dynamics, another immediate direction for future work is to determine how well one can estimate the fluid dynamics given knowledge of the locations of Lagrangian coherent structures (LCS) in the flow field.

\begin{acknowledgements}
{KM and MAH were supported by the Office of Naval Research (ONR) Award
  No. N000141211019. EF was supported by the U.S. Naval Research Laboratory (NRL) Award
  No. N0017310-2-C007. IBS was supported by ONR grant N0001412WX20083 and the NRL Base Research Program N0001412WX30002. The authors additionally acknowledge support
by the ICMAT Severo Ochoa project SEV-2011-0087.}
\end{acknowledgements}

\bibliographystyle{copernicus}
\bibliography{mapBasinBoundaries,refs}

\newpage

\begin{table}
    \begin{tabular}{lcccccc}
        \hline
        $T_c$         & 2     & 5     & 8    & 9    & 10    \\ \hline 
        Ring Pattern  & 12.99 & 5.98  & 3.45 & 3.49 & 3.66  \\ 
        Block Pattern & -     & 11.21 & -    & -    & 12.72 \\ 
        L Pattern     & -     & 30.09 & -    & -    & 30.45 \\ 
        \hline
    \end{tabular}
    \caption{Summary of the RMSE for each simulation pattern at $t=450$ with the time-invariant flow field. The RMSE for the Baseline case is 4.09.  \label{RMSEtable}}
\end{table}

\end{document}